\documentclass[twocolumn]{article}
\usepackage{bm}
\usepackage{graphicx}
\usepackage{amsmath}
\usepackage[usenames]{color}
\usepackage[utf8]{inputenc}
\usepackage{url}

\newcommand{\dd}{\;\mathrm{d}} 
\newcommand{\ff}{R}
\newcommand{\mAA}{{\text \AA}} 
\newcommand{\eq}[1]{Eq.~\ref{eq:#1}}

\newcommand{\fig}[1]{Fig.~\ref{fig:#1}}

\newcommand{\sect}[1]{\S \ref{sec:#1}}
\newcommand{\tabl}[1]{Table \ref{table:#1}}

\newcommand{\nba}[1]{}
\newcommand{\nbold}[1]{}

\newcommand{\bkgda}{\emph{\textsf{poly4b}}\;}
\newcommand{\bkgdb}{\emph{\textsf{poly5b}}\;}
\newcommand{\bkgdc}{\emph{\textsf{poly7c}}\;}
\newcommand{\bkgdd}{\emph{\textsf{expn}}}

\begin{document}

\title{Quantitative nanoparticle structures from ultrafast electron
crystallography data}

\author{
Christopher L. Farrow\\
\small Department of Applied Physics and Applied Mathematics, Columbia University,\\ 
\small New York, NY, 10027, USA \\
\and
Chong-Yu Ruan
\small Department of Physics and Astronomy, Michigan State University,\\ 
\small East Lansing, MI, 48824, USA
\and
Simon~J.~L. Billinge\\
\small Department of Applied Physics and Applied Mathematics, Columbia University,\\ 
\small New York, NY, 10027, USA \\
\small Condensed Matter Physics and Materials Science Department,\\
\small Brookhaven National Laboratory, Upton, NY 11973\\
\small \texttt sb2896@columbia.edu\\
}

\date{}
\maketitle                        

\begin{abstract}

We describe the quantitative refinement of nanoparticle structures from gold
nanoparticles probed by ultrafast electron crystallography (UEC).  We establish
the equivalence between the modified radial distribution function employed in
UEC and the atomic pair distribution function (PDF) used in x-ray and neutron
powder diffraction analysis. By leveraging PDF refinement techniques, we
demonstrate that UEC data are of sufficient quality to differentiate between
cuboctahedral, decahedral and icosahedral nanoparticle models.  Furthermore, we
identify the signatures of systematic errors that may occur during data
reduction and show that atomic positions refined from UEC are robust to these
errors. This work serves as a foundation for reliable quantitative structural
analysis of time-resolved laser-excited nanoparticle states.

\end{abstract}

\maketitle

\section{Introduction}
\label{sec:introduction}

Ultrafast electron crystallography (UEC) is a promising new approach for
studying the structure of photoexcited materials on pico and femtosecond
timescales.\cite{ruan;pnas04,ruan;nanol07}  It extends existing ultrafast
pump-probe spectroscopy techniques by providing vital structural information on
the photoexcited states, which is required to obtain a more complete
understanding of these transient states. It has long been a dream to make
``molecular movies''\cite{zewai;jpca00,zewai;aropc06,dwyer;ptrsa06} of atoms as
a material transforms from one state to another.  The use of x-rays for this
purpose, obtained from powerful synchrotron sources and in the future from hard
x-ray free electron lasers, has the advantage that the data can be modeled
quantitatively using existing x-ray crystallographic\cite{dinne;b;pdtap08} and
atomic pair distribution function\cite{egami;b;utbp03} (PDF) approaches.
However, on the experimental side, the use of electrons as a probe has a number
of advantages over synchrotron x-ray techniques.  First, the scattering power
of the electrons is higher, allowing smaller and more dilute samples to be
studied.  Second, it is easier to control the pulse structure of the
probe-pulse on femtosecond time-scales without the need for fast
choppers.\cite{gembi;jsr07}

A challenge when using electrons as a probe is that the scattering is strong,
resulting in increased multiple scattering, and the single-scattering
kinematical approximation used in most x-ray analyses may not be valid.  Recent
developments in quantitative electron
crystallography,\cite{weiri;b;ecnafsdonm06} especially with the development of
precession methods of data collection,\cite{own;aca06} show that these problems
can be overcome under certain circumstances and quantitative structure
solutions are possible using electrons.  Another challenge is structure
solution when the underlying structural correlations only extend over nanometer
dimensions, the so-called ``nanostructure problem".\cite{billi;s07}  Here we
explore whether methods developed for quantitative structure
solution\cite{juhas;n06} and
refinement\cite{proff;jac99,farro;jpcm07,tucke;jpcm07} of x-ray or neutron
derived PDFs can be extended to UEC data from nanoparticles.  The full
potential of UEC can be realized only when the quantitative reliability of
structures determined from the data is established.  We investigate this issue
here for the case of data from metallic nanoparticles tethered to a surface.

There is a large literature on the structure solution of small molecules in the
gas phase from electron diffraction data.\cite{iwasa;ac64}  However, the UEC
experiments of interest are on samples in the solid or liquid phase where the
validity of the kinematical approximation is less clear.  Quantitative
structure solutions from electron crystallography based on kinematic scattering
are possible using very thin crystals,\cite{weiri;n96, wagne;jpcb99} weakly
scattering specimens\cite{moss;jpsb82, wu;aca02} and nanoparticles containing
just 10-20~unit cells.\cite{huang;nm08} In these cases, the probability of a
multiple-scattering event is sufficiently small that it can be treated as a
minor perturbation on the single-scattering signal, and the data can be
properly corrected for a quantitative structure solution.\cite{dorset;sec95}
Here we investigate whether UEC data from nanoparticles tethered to a surface
can be treated in this way.  There are powerful tools emerging for the study of
nano-scale structures from ensembles of nanoparticles using x-ray and neutron
diffraction
data.\cite{billi;s07,juhas;n06,mcgre;jpcm01,tucke;jac01a,masad;prb07,neder;jpcm05}
These tools can also be applied and extended for studying UEC data if the
quantitative reliability of the data in the kinematical limit can be
established.

In this paper we show that it is possible to obtain UEC data of sufficient
quality for quantitative structure modeling. We investigate the robustness of
structural solutions with respect to uncontrolled aspects of the data reduction
such as the use of empirical methods for removing incoherent backgrounds and
determining an effective electron-form-factor that considers the absorption by
the substrate. We elucidate potential sources of error in the data reduction
procedure, and indicate the signature those errors.  Additionally, we show the
relationship between the correlation functions traditionally used in gas-phase
electron diffraction, and those used in the study of nanostructure with x-rays
and neutrons, bringing together these two fields of study. These results place
the UEC method on a more sound footing with respect to the study of ensembles
of nanoparticles, opening the door to reliable quantitative structure solution
of transient states, and therefore physically meaningful molecular movies of
these materials.

\section{UEC with nanoparticles}
\label{sec:experiment}

The photochemical processes of bond breaking and forming can be resolved using
femtosecond laser techniques in a pump-probe configuration, in which the first
laser pulse (pump) is used as a trigger to initiate the chemical reactions and
the ensuing probe laser pulse serves to monitor the changes via spectroscopic
responses.\cite{zewai;jpca00} By replacing the probe laser pulse with a
laser-triggered ultrashort electron pulse, the transient structural dynamics
can be directly probed.  This ultrafast electron diffraction
(UED)\cite{srini;hca03}  method was shown to be especially advantageous in the
studies of optically dark processes, such as nonequilibrium structural
transformation,\cite{ruan;pnas01} and radiationless relaxation channels for the
photo-excited state.\cite{srini;s05}

The UEC development extends UED to the condensed phase, and has been applied to
study the collected excitations involving photons, electrons and phonons from
the interfacial water bilayers to
superconductors.\cite{ruan;s04,baum;s07,gedik;s07} Whereas the periodic orders
from the crystalline samples offer higher signal-to-noise ratio (SNR), and
potentially, higher spatiotemporal resolutions, the complexity of the
anisotropic scattering in the dynamical scattering regime poses significant
challenges in quantitatively assessing all the possible details of the relevant
atomic dynamics that the higher SNR can now provide.

\begin{figure}
\includegraphics[width=0.9\columnwidth]{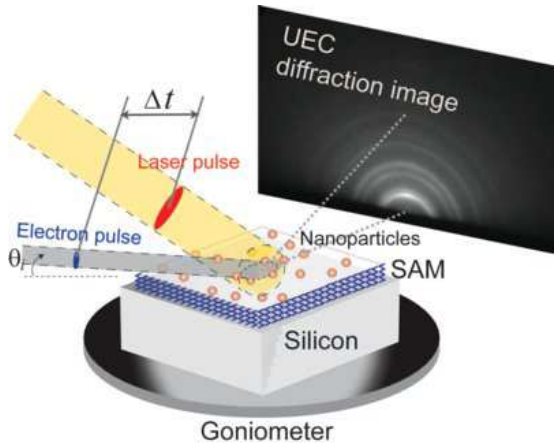}
\caption{\label{fig:UEC}
\nba{Purpose: Show the UEC exp geometry.}
UEC for nanoparticles experimental setup, implementing the
laser-pump-electron-probe configuration.  The electron pulse is delayed
relative to the 'pumping' laser pulse in a time-sequence ($\Delta t$) to
examine the structural dynamics.  The incident angle of the electron beam is
kept low, typically 1-2$^\circ$, to achieve high sensitivity to the
nanoparticles, which are dispersed on a self-assembled molecular monolayer atop
a supporting silicon substrate, to produce powder-like diffraction patterns.
}
\end{figure}

Recently, UEC has been applied to investigate nanoparticles that are highly
dispersed on a surface to produce powder-like patterns for size-scale down to
20~\AA.\cite{ruan;nanol07} To suppress the background signals from the
supporting substrate, soft-anchoring of the nanoparticles using self-assembled
molecular (SAM) layer on silicon surface is employed. The implementation of
this specific UEC for studying nanoparticles (as shown in \fig{UEC}) affords a
low coverage to ensure the scattering from isolated particles can be recorded
unobstructively, and provides the needed chemical and physical isolation of
nanoparticles under repetitive time-resolved studies.\cite{ruan;mm09}

Even though a relatively clean coherent diffraction pattern for nanoparticles
can be obtained this way by rendering the scattering from the substrate largely
diffusive, it is still unclear how the residual effects caused by the
uncontrolled diffusive background and the surface absorption affect the
quantitative retrieval of structural functions. The empirically derived
structural functions from UEC\cite{ruan;mm09} will be refined using standard
x-ray PDF analysis to examine the self-consistency and robustness of UEC data
for quantitative structural analysis.

\section{Correlation functions}
\label{sec:functions}

In this section we discuss the relationship between the modified radial
distribution function (mRDF) obtained from UEC experiments with the PDF
obtained from x-ray and neutron measurements. Historically, the mRDF has been applied in gas-phase
electron diffraction to study atom-pair correlations in small molecules
containing a few to tens of atoms. In UEC, the mRDF is used to study the change
in atomic arrangements in excited-state bulk- and
nano-structures.\cite{ruan;nanol07,raman;prl08}

The mRDF, $\ff(r)$, is defined as the truncated Fourier transform of the
molecular scattering intensity,\cite{hargi;b;saogped88}
\begin{equation}
\label{eq:mrdf}
\ff(r) = \int_0^{s_{max}} sM(s) \sin(sr) \dd s,
\end{equation}
where $s$ is the magnitude of the scattering momentum transfer and $s_{max}$ is
the maximum determined value of $s$.  The molecular scattering intensity,
$M(s)$, contains the structural information extracted from the scattered
electron intensity. The mRDF and $sM(s)$ contain the same structural
information, and it is the $sM(s)$ curve that is typically used in structural
modeling.\cite{hedbe;ac64,iwasa;ac64}

The PDF, $G(r)$, is used in x-ray and neutron investigations of liquids, and
amorphous and nanostructured materials.\cite{egami;b;utbp03} The PDF is the
truncated Fourier transform of the reduced total scattering structure function,
$F(Q)$.\cite{farro;aca09}
\begin{equation}
\label{eq:pdf}
G(r) = \frac{2}{\pi} \int_{Q_{min}}^{Q_{max}} F(Q) \sin(Qr)\dd Q,
\end{equation}
where $Q$ is the magnitude of the scattering momentum
transfer\cite{warre;b;xd90} and $Q_{min}$ and $Q_{max}$ are the measured
$Q$-extrema. For elastic scattering, $Q= (4 \pi/\lambda) \sin(\theta)$, where
$\lambda$ is the wavelength of the probe, and $2\theta$ is the scattering
angle.  Note, the scattering angle in the gas phase electron diffraction
literature is generally denoted $\theta$, rather than $2\theta$.  Notation
aside, $Q$ and $s$ are equivalent:
\begin{equation}
Q \equiv s.
\end{equation}
In deference to the respective literatures, and to help differentiate between
the two sets of functions, we will use $s$ when referring to UEC and $Q$ in
reference to x-ray and neutron diffraction.

As we will show in the next section, $F(Q)$ and $sM(s)$ are effectively the
same, where the only differences come about due to experimental effects and
uncertainties in the data reduction process. These being equal, the important
difference between the mRDF and PDF is the lower integration limit in their
respective definitions.\cite{farro;aca09}

There is always a finite lower bound on the measured scattering momentum ($s$
or $Q$) due to experiment geometry and physical limitations of measurement
equipment. It is the practice in gas-phase electron diffraction and UEC to
compensate for this missing scattering intensity by using a presumed structural
model.\cite{hargi;b;saogped88,ruan;nanol07} In gas-phase electron diffraction
the missing intensity below $s_{min}$ is extrapolated to $s=0$ using the
model, and in UEC the same effect is achieved by adding a baseline to the
transformed real-space signal.  This works well when the system under study is
made up of individual small molecules in a dilute gas phase.  However, it
becomes highly problematic in condensed systems.  In this case, the convention
in x-ray and neutron diffraction is to ignore any scattering in the low-$Q$
limit.

When the Fourier transform includes the intensity extending to $s=Q=0$,
one obtains the mRDF, which is always positive (apart from termination effects
due to a finite $s_{max}$). However, if the small angle scattering is excluded,
one obtains the PDF, which is a function that oscillates around zero,
sitting on top of a negative baseline.\cite{farro;aca09} In either case, with
the same minimum and maximum momentum transfer, the same structural information
can be obtained from the mRDF or the PDF.  We use the PDF in this study in
order to avoid the extra step of compensating for the missing scattering
information.

\section{Data reduction in detail}
\label{sec:reduction}

In this section we will describe how $sM(s)$ is obtained in practice in order
to examine the systematic errors that may affect the extraction of structural
information from the data. We will first discuss how the total scattering
structure function is typically obtained in x-ray and neutron diffraction. This
process has an established theoretical foundation,\cite{egami;b;utbp03} and
will serve as a basis of comparison with the UEC data reduction procedure.

\subsection{X-ray and neutron diffraction data reduction}

The reduced total scattering structure function, $F(Q)$, is defined in terms of
the total scattering structure function, $S(Q)$, as $F(Q) = Q(S(Q)-1)$. The
structure function contains the discrete coherent singly scattered information
available in the raw diffraction intensity data.  It is defined according
to\cite{farro;aca09}
\begin{equation}
\label{eq:sofq}
S(Q) = \frac{I_c(Q)}{N \langle f\rangle^2} - \frac{ \langle (f - \langle f
\rangle)^2 \rangle}{\langle f \rangle^2},
\end{equation}
which gives
\begin{equation}
\label{eq:sofqm1}
\begin{split}
S(Q) - 1 =& \frac{I_c(Q) - N \langle f^2\rangle}{N\langle f\rangle^2}\\
=& \frac{I_d(Q)}{N\langle f\rangle^2},
\end{split}
\end{equation}
where $f$ is the $Q$-dependent x-ray scattering factor or $Q$-independent
neutron scattering length as appropriate and $\langle \ldots \rangle$
represents an average over all atoms in the sample.  In this equation, $I_c(Q)$
is the coherent single-scattered intensity per atom and $I_d(Q)$ is the
discrete coherent scattering intensity, which excludes the self-scattering,
$N\langle f^2\rangle$.\cite{farro;aca09}  The coherent scattering intensity is
obtained from the measured intensity by removing parasitic scattering (e.g.,
from sample environments), incoherent and multiple scattering contributions,
and correcting for experimental effects such as absorption, detector
efficiencies, detector dead-time and so on.\cite{egami;b;utbp03} The resulting
corrected measured intensity is normalized by the incident flux to obtain
$I_c(Q)$. The self-scattering, $N\langle f^2\rangle$, and normalization,
$N\langle f\rangle^2$, terms are calculated from the known composition of the
sample using tabulated values of $f$.

As evident in \eq{sofq}, to obtain $S(Q) - 1$ from $I_c(Q)$ we subtract the
self-scattering, $N\langle f^2\rangle$, which has no atom-pair correlation
information, and divide by $N\langle f\rangle^2$.  As a result, $S(Q) - 1$
oscillates around zero, and asymptotically approaches it at high~$Q$ as the
coherence of the scattering is lost.  If the experimental effects are removed
correctly, the resulting $F(Q)$ and $G(r)$ are directly related to, and can be
calculated from, structural models.\cite{farro;aca09} The corrections are well
controlled in most cases and refinements of structural models result in reduced
$\chi^2$ values that approach unity in the best cases.  Some uncertainty in the
corrections can be tolerated because they are mostly long-wavelength in nature
(for example, the absorption correction), which results in aberrations in
$G(r)$ at very low-$r$ below any physically meaningful region of the
PDF.\cite{peter;jac00}

\subsection{UEC data reduction}
\label{sec:uecreduction}

In practice, $M(s)$ is obtained from the raw experimental intensity, $I(s)$, by
an empirical approach.\cite{ruan;mm09} This approach is necessary because the
reflection geometry employed to effectively sample the surface-dispersed
nanoparticles results in a shadowing effect, i.e. the diffraction intensities
collected at small exit angles are strongly attenuated due to surface
absorption.  In practice, due to the random nature of the surface roughness,
the attenuation factor is not known precisely.  It is not immediately clear
that the empirical method results in quantitatively reliable mRDFs, as
demonstrated in a more simple transmission geometry in which the tabulated
atomic scattering form factor can be used directly to yield quantitative
structure solutions of molecular systems.\cite{srini;hca03}

\begin{figure}
\includegraphics[width=\columnwidth]{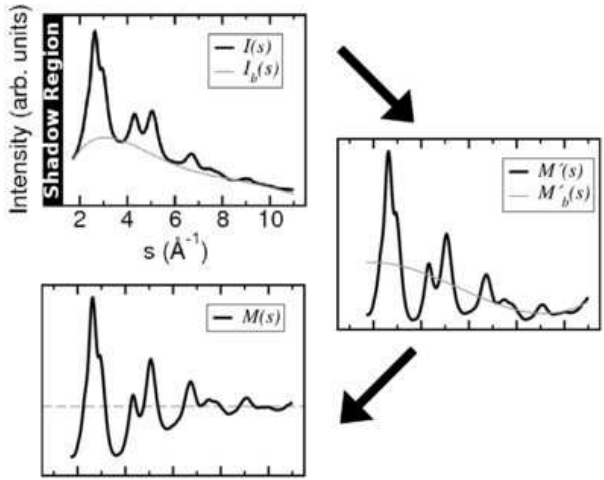}
\caption{\label{fig:datareduction}
The UEC data reduction procedure showing the reduction of $I(s)$ to
$M(s)$.
}
\end{figure}

The UEC data reduction procedure is demonstrated in \fig{datareduction}. First,
an empirical background is estimated by fitting a slowly varying function, such
as a low order polynomial, through the center or base of the raw intensity.
This background is subtracted from the raw intensity.  The
background-subtracted data are then divided by the same background function.
We denote this estimated background as $I_b(s)$ and can write the resulting
intermediate intensity,
$M'(s)$, as
\begin{equation}
\label{eq:mps}
M'(s) = \frac{I(s) - I_b(s)}{I_b(s)}.
\end{equation}
\nba{Chong-Yu has a great figure from his review.}
\nba{how is this done in practice?  "Fit" implies  minimization of some
target function, but most chi2's I can think of would fit the center of mass,
not the baseline (in fact a baseline fitting chi2 this is something we would
love in Rietveld).  Let's think about expanding to be more explicit.}
\nba{CY's comment: The fit uses nonnegativeness as the target function, which is obtained by summing the negative components in the residue and turning it into a positive number.}
\nba{clf: Are we going to include this? It is useful information, but not
relevant to this section.}

The next step in processing the data is to adjust for any offset in the
initial background. This secondary background, $M_b'(s)$, captures the
slowly-varying components of $M'(s)$, and as a result the remaining signal
oscillates around zero. The result is the molecular scattering intensity,
\begin{equation}
\label{eq:ms}
M(s) = M'(s) - M_b'(s).
\end{equation}

\subsection{Practical relationship between the molecular scattering and the
total scattering structure function}
\label{sec:reductionrelation}

In order to understand how $M(s)$ relates to $S(Q)$, we consider an ideal case;
scattering from a monatomic system with no undesirable experimental effects
influencing the data. If measured without an offset, $I_b$ captures only the
slowly varying self-scattering from the (normalized coherent) scattering
intensity.  This self scattering is equal to $N\langle f\rangle^2$ for a
monatomic system.  Thus, $M'(s) = (I_c(s) - N\langle f^2\rangle)/N\langle
f\rangle^2 = I_d(s)/N\langle f\rangle^2 = S(s) - 1$.  Since this background is
measured perfectly,  $M_b' = 0$, and so $M(s) = M'(s) = S(s) - 1$.

In reality, the data are not ideal; $I_b$ must correct for experimental
effects, and $M_b'$ must somehow compensate for the accuracy of $I_b$.  More
formally, we write the raw intensity as the coherent intensity modified by
multiplicative term that captures multiplicative experimental effects (e.g.,
absorption correction) and an additive term representing the incoherent
background and additive experimental effects (e.g., substrate scattering,
inelastic scattering)
\begin{equation}
I(s) = \sigma(s)I_c(s) + \nu(s).
\end{equation}

The background intensity, $I_b(s)$, estimates the modified self-scattering
$\sigma(s) N\langle f^2\rangle$, the incoherent scattering background, and the
undesirable additive effects, $\nu(s)$.  The goal in subtracting $I_b(s)$ is to
remove the self-scattering and additive effects from the intensity.  This
background is typically estimated using a low-order polynomial, which can
approximate the experimentally modified $N\langle f^2\rangle$ with relative
accuracy.\cite{doyle;aca68} The estimate is subject to error, however, which
may be due to components of $\nu(s)$ that are poorly estimated or due to an
intentional offset of the estimated background curve. We denote this error by
$-\epsilon(s)$. Thus, the first stage of the UEC data reduction results in
\begin{equation}
\label{eq:estmps}
\begin{split}
M'(s) &= \frac{\sigma(s)I_c(s) + \nu(s) - (\sigma(s)N\langle f^2\rangle + \nu(s) - \epsilon(s)))}
{\sigma(s)N\langle f^2\rangle + \nu(s) - \epsilon(s)}\\
&= \frac{\sigma(s)(I_c(s) - N\langle f^2\rangle)}{\sigma(s)N\langle f^2\rangle + \nu(s) -
\epsilon(s)} \\
&     + \frac{\epsilon(s)}{\sigma(s)N\langle f^2\rangle + \nu(s) - \epsilon(s)}\\
&= \frac{\sigma(s)(I_c(s) - N\langle f^2\rangle)}{\sigma(s)N\langle f^2\rangle + \nu(s) - \epsilon(s)}
+\gamma(s).
\end{split}
\end{equation}
Note that $\gamma (s)$ and the denominator in \eq{estmps} are slowly
varying functions of $s$.

The next step is to eliminate the effects of $\gamma(s)$, which result from
the offset in the estimate of $I_b(s)$. Since $\gamma(s)$ is slowly varying it
can be fit with a low-order polynomial. With $\gamma(s)$ eliminated, the result
will oscillate around zero. There will be some error in this estimate as well,
which we denote $-\beta(s)$.  Thus,
\begin{equation}
M_b'(s) = \gamma(s) - \beta(s).
\end{equation}
Using this expression for $M_b'(s)$ gives
\begin{equation}
M(s) = \frac{\sigma(s)(I_c(s) - N\langle f^2\rangle)} {\sigma(s)N\langle f^2\rangle + \nu(s) -
\epsilon(s)} + \beta(s).
\end{equation}
This is very near to the form of \eq{sofqm1}. To get this in the desired form,
we write $I_c(s) - N\langle f^2\rangle$ = $(S(s)-1) N\langle f \rangle^2$ (see
\eq{sofqm1}) so that
\begin{equation}
\label{eq:msvss}
M(s) = \alpha(s)(S(s) - 1) + \beta(s),
\end{equation}
where
\begin{equation}
\label{eq:alpha}
\alpha(s) = \frac{\sigma(s) N\langle f \rangle^2}
            {\sigma(s) N\langle f^2\rangle + \nu(s) - \epsilon(s)},
\end{equation}
and we note that $\alpha(s)$ is also slowly varying.  Now, understanding the
effective difference between $M(s)$ and $S(s) -1$ reduces to understanding the
effects $\alpha(s)$ and $\beta(s)$ have on the measured correlation function.

First we consider $\beta(s)$. By construction, $\beta(s)$ is small, and
contains only slowly varying components.  As mentioned above, this means that
$\beta(s)$ contributes only to the low-$r$ region of the correlation function,
below the physically interesting region.\cite{peter;jac00} If the experimental
effects are not slowly varying, then the Fourier transform of $\beta(s)$ will
leak into the structurally relevant portion of the correlation function. The
effect is unpredictable, but should be small.

The multiplicative term, $\alpha(s)$, is likely to have a more significant
effect on the correlation function.  In the best case, $\alpha(s)$ is constant
and it scales the peaks of the correlation function uniformly, which does not
obscure the structural information. The only aspect of $\alpha(s)$ that is
controlled by the data reduction is how quickly it oscillates. This is
controled by how well the modified self-scattering is fit by the intensity
background (manifest in $\nu(s)$ and $\epsilon(s)$). Depending on how well
these factors are estimated, $\alpha(s)$ may destort the peak profile of the
correlation function. At worst, this could result in a measurable peak shift,
which would complicate the extraction of reliable structure parameters.

\section{Methods}
\label{sec:methods}

Data have been collected on size-selected gold nanoparticles with diameter of
approximately 20~\AA~using the UEC experimental setup described in
\sect{experiment}. Further details on the sample preparation and data
collection can be found elsewhere.\cite{ruan;nanol07}  To gain quantitative
information about the morphology of gold nanoparticles we modeled PDFs from
experimental UEC data using three structure models: a 309-atom cuboctahedron
(diameter 22.3~\AA), a 309-atom Mackay icosahedron\cite{macka;ac62} (diameter
23.1~\AA) and a 181-atom decahedron\cite{ino;jpsj66} (diameter 23.8~\AA).  These
gold cluster types have been shown to exist at a similar size\cite{cerve;jac03}
and have been the subject of a previous UEC study.\cite{ruan;mm09} We also
fit these nanoparticle models with one fewer and one additional shell. The
results in those fits were considerably worse than for these nanoparticle
sizes, and will not be reported.

\begin{figure}
\begin{tabular}{ccc}
\centering
\includegraphics[width=0.33\columnwidth]{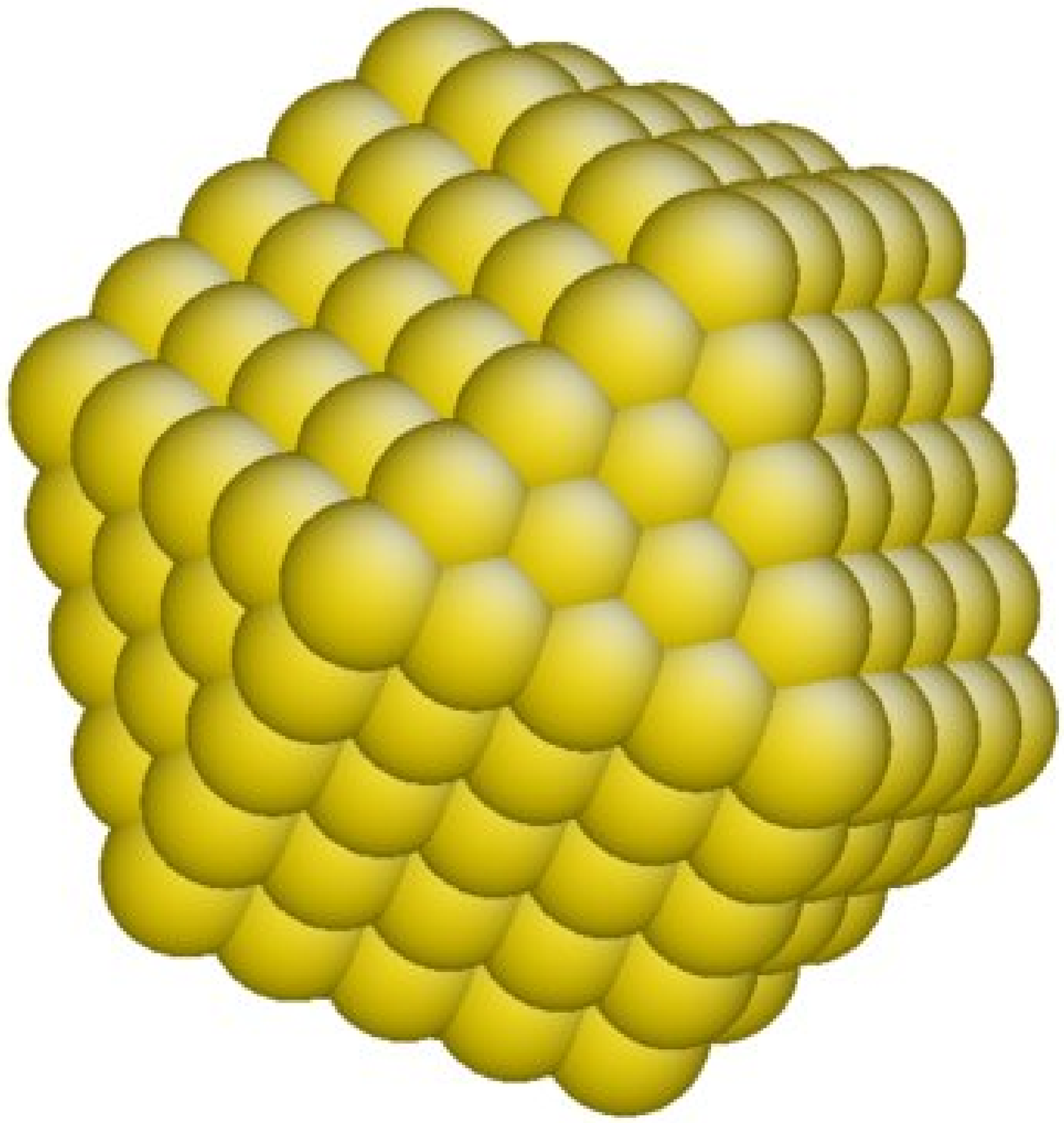} &
\includegraphics[width=0.33\columnwidth]{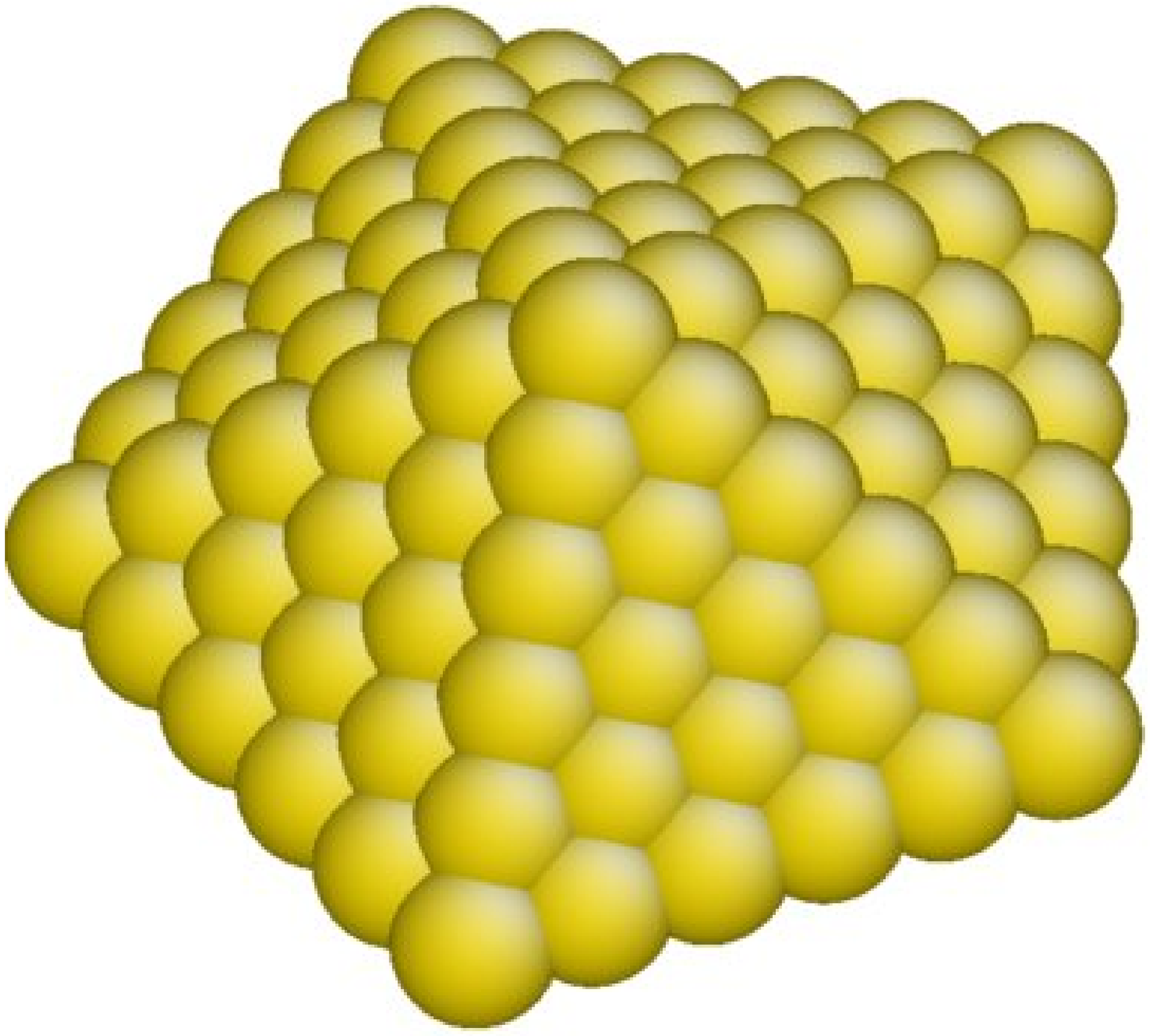} &
\includegraphics[width=0.33\columnwidth]{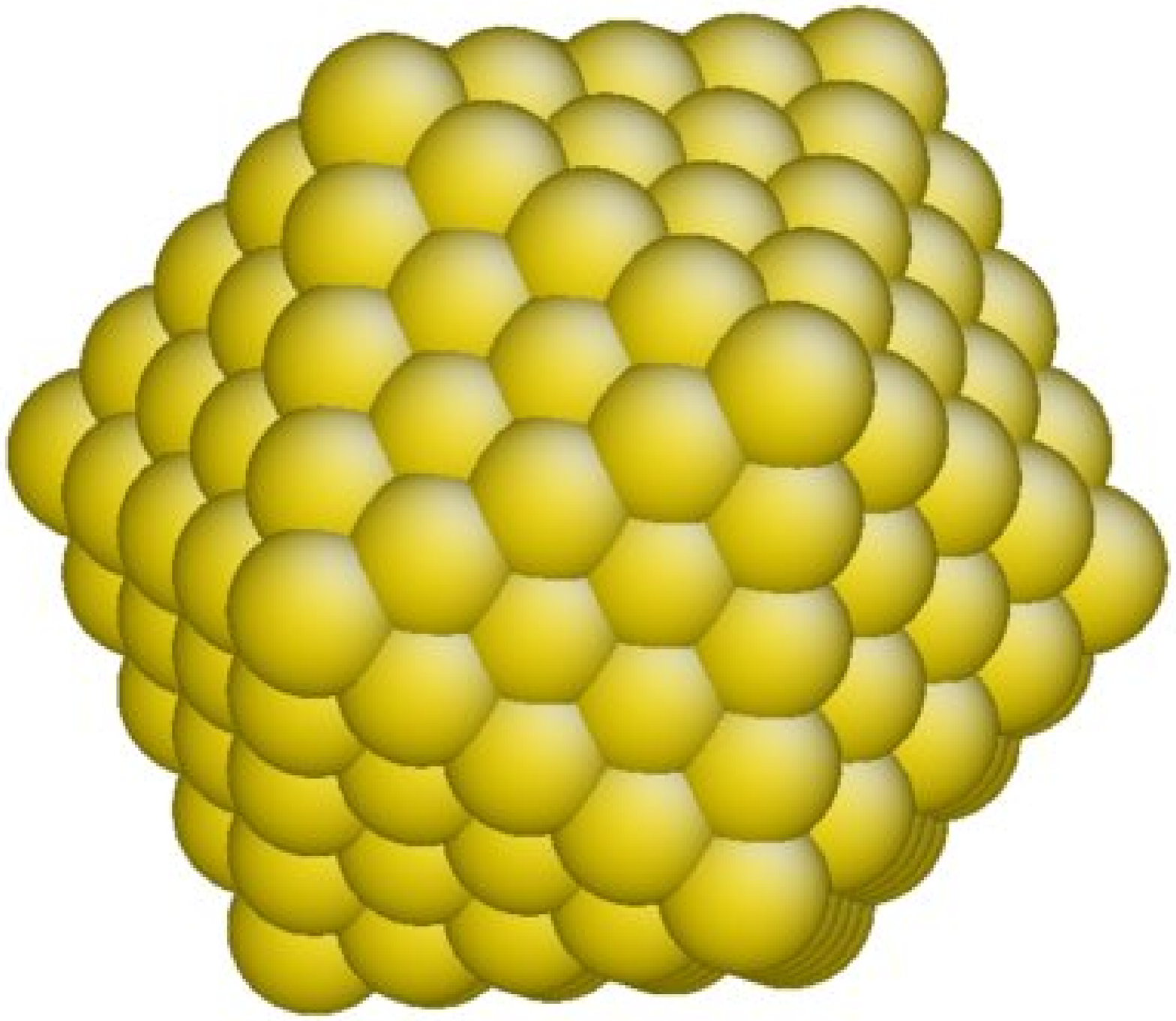} \\
\multicolumn{3}{l}{\includegraphics[angle=-90,width=\columnwidth]{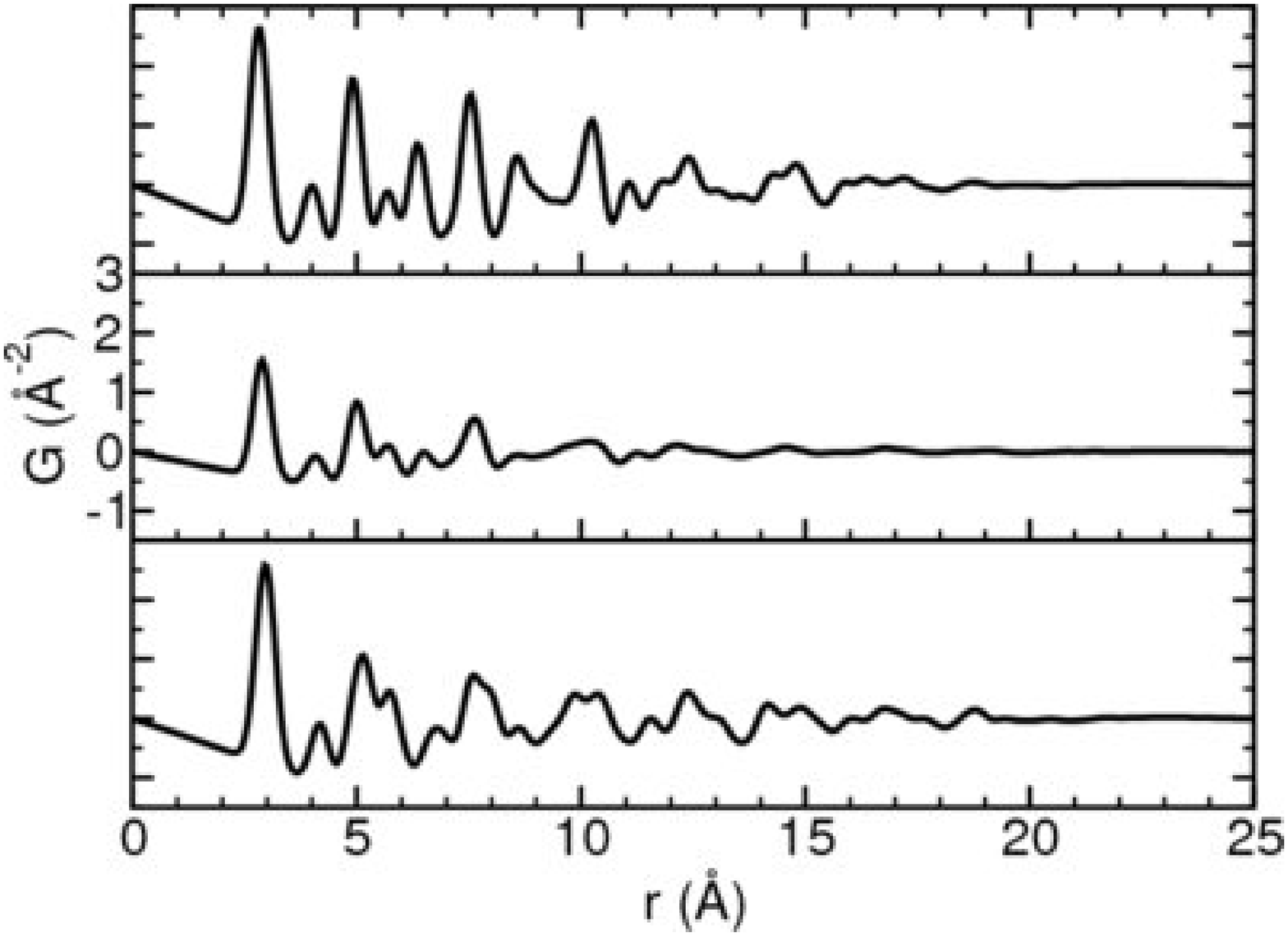}}
\end{tabular}
\caption{\label{fig:models}
\nba{Purpose: Show the nanoparticle models we used.}
PDFs from likely models for $\sim20~\mAA$ gold nanoparticles.
Top: Cuboctahedral, decahedral and icosahedral models.
Bottom: PDFs calculated from the cuboctahedral (top), decahedral (middle) and
icosahedral (bottom) models.
}
\end{figure}

The structure models are shown in \fig{models} along with the theoretical PDF
from each model.\footnote{Figures created with the AtomEye structure
viewer.\cite{li;msmse03}} As can be seen in that figure, the models give PDFs
that are very different past 5~\AA, which reflects their topographical
differences.  All nanoparticles were modeled with rigid atomic arrangements and
uniform isotropic atomic displacement parameters (ADPs).

In all refinements the nanoparticle models were allowed to expand or contract
isotropically keeping the number and relative positions of atoms in each model
fixed.  The samples are nominally mono-disperse, so to avoid complicating the
analysis we have not attempted to model size distributions. Furthermore, we
have not considered strain effects besides a simple isotropic expansion of the
models.  

We calculate $F(Q)$ from a structure model using the Debye
equation,\cite{debye;ap15,farro;jpcm07} modified such that the
pair-contributions are attenuated by a Debye-Waller factor, and this is
transformed to $G(r)$ according to \eq{pdf}. The Debye-Waller factor is derived
from the ADPs and a refinable vibrational correlation
term.\cite{jeong;jpc99,jeong;prb03} A scale factor was refined to account for
the undetermined scale of the raw intensity.  Resolution
factors\cite{toby;aca92} that broaden and dampen the PDF peaks were also
refined to simulate the effects of the finite $s$-resolution of the
measurement.  The refinements were performed using a home-written least squares
regression algorithm based on the \textsc{PDFfit2} program.\cite{farro;jpcm07}

We distinguish the fits by their agreement with the data. This is quantified
with the weighted residual of the PDF, defined as
\begin{equation}
\label{eq:rw}
R_w = \sqrt{ \frac{\sum_i \omega(r_i)(G(r_i) - G_{m}(r_i))^2}{\sum_i
\omega(r_i)G(r_i)^2} },
\end{equation}
where $G$ denotes the experimental PDF and $G_{m}$ is the PDF calculated from a
structure model.  For this study the weights $\omega(r_i)$ are equal for every
point and therefore cancel in $R_w$.  The $R_w$ factor was calculated only over
the physical portion of the fit range, which we estimate to be between 2.2 and
20~\AA.

To test the robustness of the data reduction and refinement procedure, we have
also performed refinements on the PDF produced from four different data
reduction protocols and compared the results. These protocols differ in how the
intensity background, $I_b(s)$, is determined. The different protocols produce
different $s$-dependent scaling factors, $\alpha(s)$, and ultimately different
PDFs.  For two reduction procedures, $I_b(s)$ is determined by fitting along
the bottom of $I(s)$ using 4th and 5th degree polynomials (denoted \bkgda~and
\bkgdb, respectively), the latter of which is the standard procedure.
\cite{ruan;pnas04} The other two procedures fit $I_b(s)$ through the center of
$I(s)$ using a 7th degree polynomial (denoted \bkgdc) and an exponential function of the form $\sqrt{s-s_0}\exp(-s^{0.75})$ (denoted \bkgdd) that is
empirically found to well describe the shadowing effect and decay of the
scattering signal.

\section{Results}
\label{sec:results}

\subsection{Quantitative Structure Modeling}

To determine whether the UEC data are of sufficient quality to differentiate,
quantitatively, between structural models, we fit cuboctahedral, decahedral and
icosahedral nanoparticle models to UEC PDF data produced using the standard
procedure (\bkgdb) as described in \sect{methods}.

\begin{figure}
\includegraphics[angle=-90,width=\columnwidth]{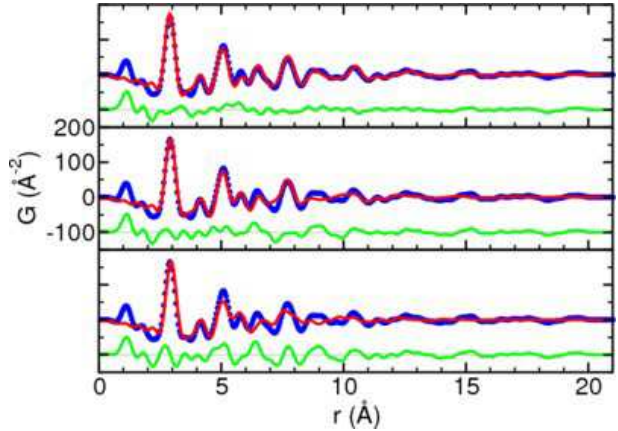}
\caption{\label{fig:stdfits}
\nba{Purpose: Show fits to the standard data using different models.}
Fits to the standard data. \bkgdb~data (blue dots), fit (red line) and
difference (green line, offset).
(Top) Cuboctahedral model.
(Middle) Decahedral model.
(Bottom) Icosahedral model.
Note the comparative quality of the cuboctahedral fit above 5~\AA~and generally
poor fit to the third and fourth peaks.
}
\end{figure}

Among the models used in these refinements, the cuboctahedral model stands out
as the best, resulting in a $R_w$ value of 0.23, compared to 0.30 for the
decahedral model and 0.41 for the icosahedral model.  These fits fits are shown
in \fig{stdfits}.  None of the structure models reproduce the third and fourth
peaks of the PDF well. It is possible that this misfit is due to errors
introduced during the data reduction. We explore this possibility below. The
structural results from the cuboctahedral model are shown in \tabl{cubresults}.

\subsection{Robustness}

\begin{figure}
\includegraphics[angle=-90,width=\columnwidth]{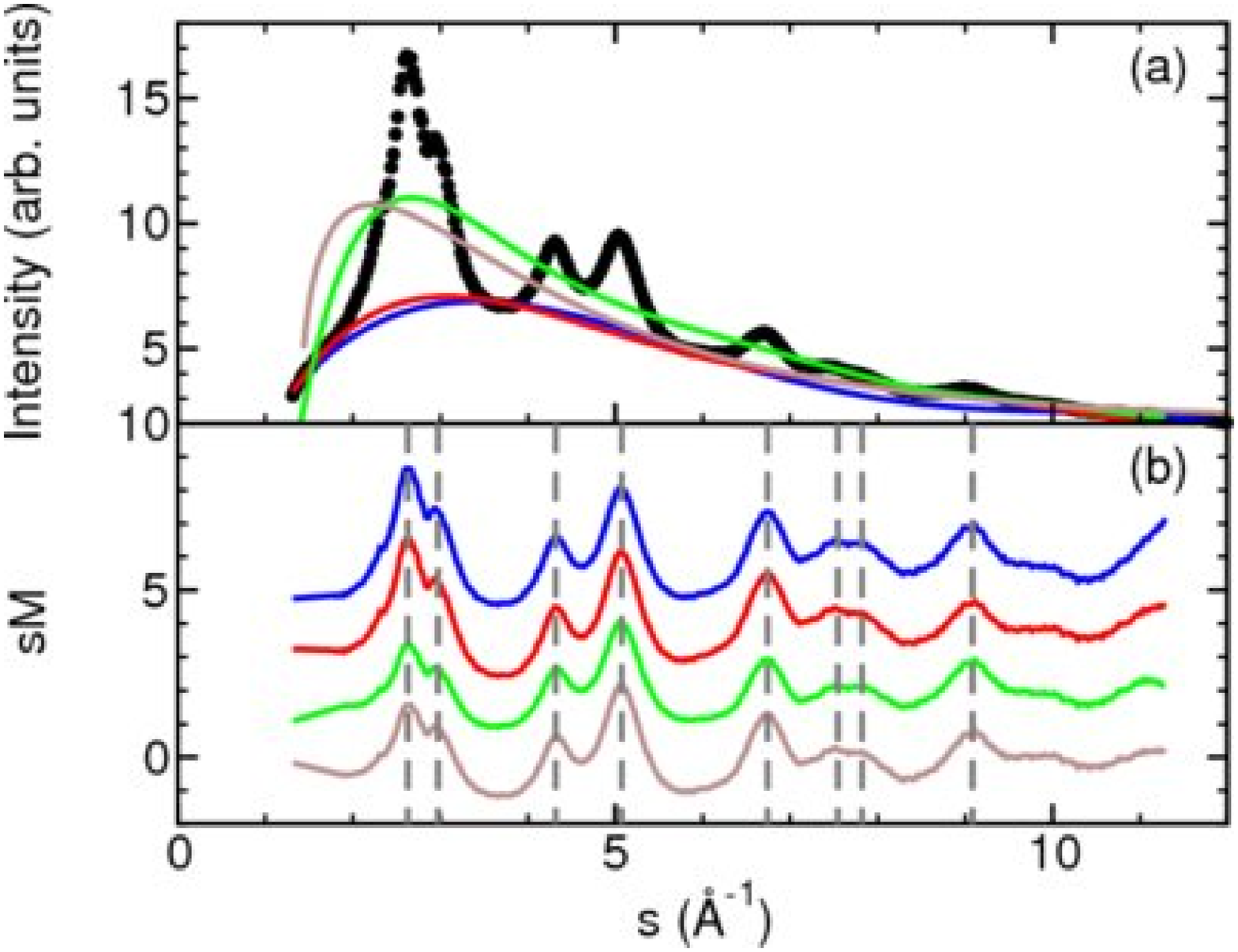}
\includegraphics[angle=-90,width=\columnwidth]{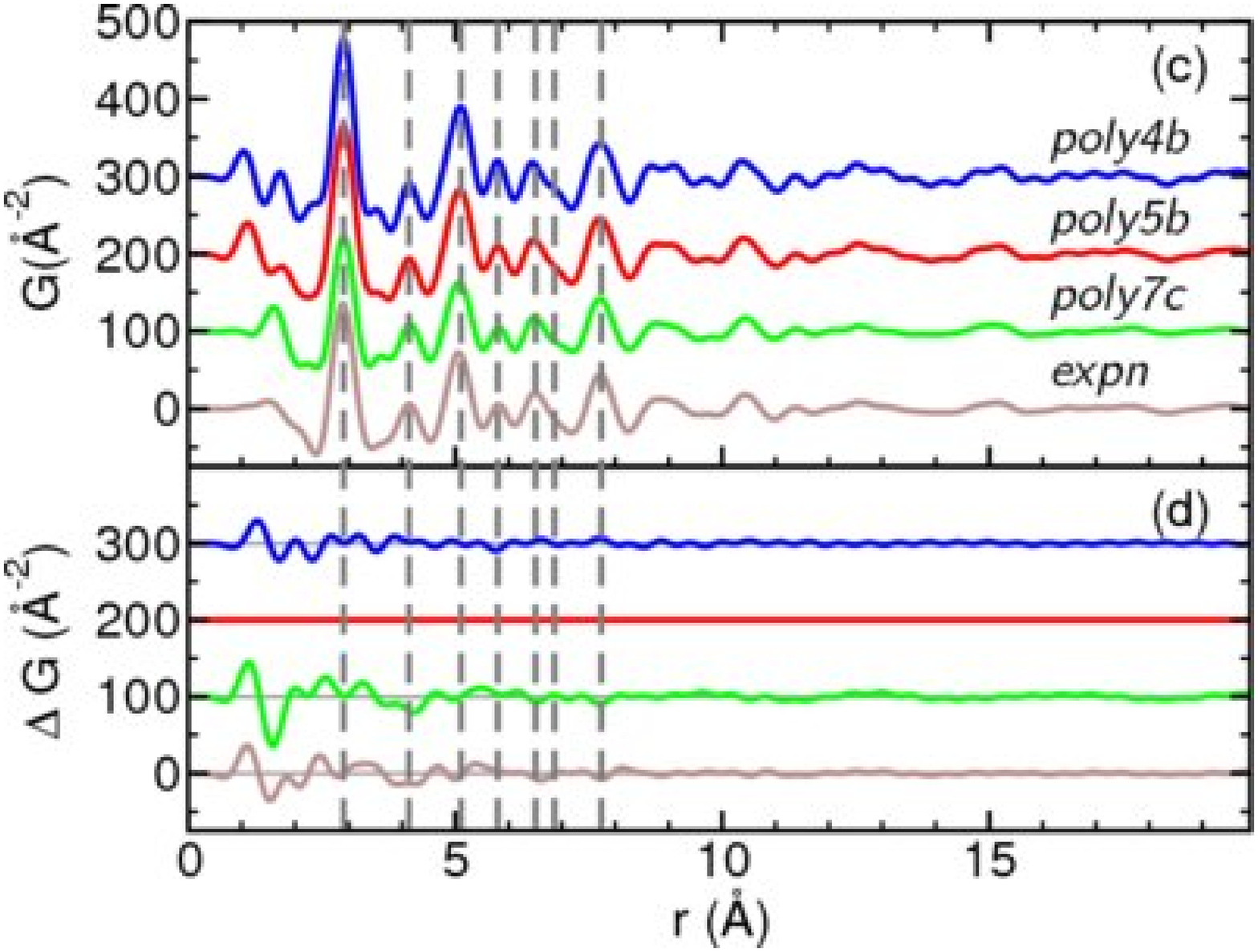}
\caption{\label{fig:reductions}
\nba{Purpose: Demonstrate that the data reduction produces very similar $sM(s)$
and PDF curves.}
Similarity between structure functions from various data reduction procedures.
Blue line: \bkgda.
Red line: \bkgdb.
Green line: \bkgdc.
Brown line: \bkgdd.
See text.
Note the similarity of the major profile features, as indicated by the dashed
lines.
(a) $I_b(s)$ for each protocol through the raw intensity (black circles).
(b) $sM(s)$ for different reductions.
(c) PDFs from the different reductions.
(d) PDF differences taken from the \bkgdb~PDF. All PDFs are
scaled to have the same first-peak amplitude. The structural region is $r >
2.2$~\AA.
}
\end{figure}

Given the empirical nature of the methods for producing the PDF from the
diffraction intensity, it is worthwhile to investigate the robustness of
refinement results based on the different procedures. This will give insight
into the influence of the data reduction procedure on the PDF, and will give us
a quantitative measure of the systematic bias that may be introduced during
data reduction.  We have performed refinements of the cuboctahedral model
described above to the PDF produced from the four different data reduction
procedures described in \sect{methods} and demonstrated in \fig{reductions}.

As can be seen in \fig{reductions}, the data reduction procedures all produce
$sM(s)$ and PDF curves similar to each other. The major features of these
curves (indicated by the dashed lines in the figure) are preserved using each
procedure. The largest differences among the PDFs appear in the peak widths and
amplitude ratios. Peak asymmetry is hard to detect given that the PDF sits on a
negative baseline, and the peaks are therefore not Gaussian.  The bottom panel
of \fig{reductions} shows the difference between the PDF from the
\bkgdb~background (the standard procedure) and the other PDFs, scaled so that
the heights of the first peaks are identical. The pre-peak signal is rather
different in each signal, indicative of the additive errors, $\beta(s)$, in
$sM(s)$ resulting from the data reduction procedures.

\begin{figure}
\includegraphics[angle=-90,width=\columnwidth]{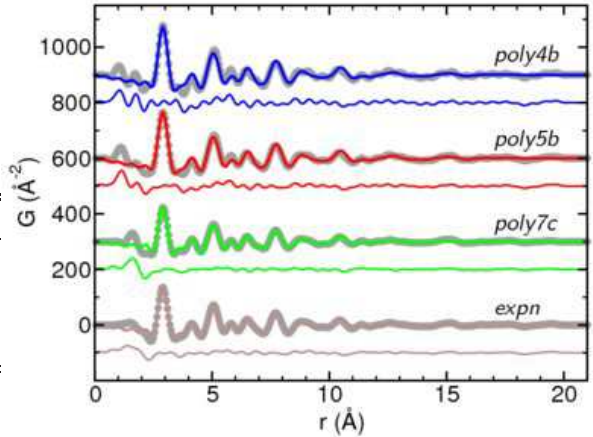}
\caption{\label{fig:allfitscub}
\nba{Purpose: Show that fits to the data are roughly the same quality.}
Fits using the cuboctahedral model to PDFs produced with different data
reduction protocols. (See \fig{reductions} for color key.)
PDF fits (solid lines), data (gray circles) and difference between data and
fit (solid lines, offset below).
}
\end{figure}

Fits to the data were performed using the cuboctahedral model as described in
\sect{methods}.  The cuboctahedral model fits can be seen in \fig{allfitscub}
and are summarized in \tabl{cubresults}. In the table, $a$ is the effective FCC
lattice constant derived from the interior nearest neighbor distance.  As seen
from the figure and table, all fits are of variable quality, with a markedly
improved $R_w$ value for the exponential background. It can be seen that this
fit reproduces the fourth peak significantly better than the other fits.

From \tabl{cubresults} we see that the measured peak width parameter
($U_{iso}$) is not robust with respect to the data reduction procedure. This
indicates that the at least some of the data suffer from peak profile
distortions due to the data reduction protocol. Considering the ideal data
reduction scenerio discussed in \sect{reductionrelation}, where the ideal
intensity background perfectly matches the modified self-scattering through the
center of the raw intensity, we believe that protocols \bkgdc~and \bkgdd, which
estimate the background through the center of the raw intensity, introduce less
error into the correlation function.  This conclusion is supported by the
superior low-$r$ fit agreement to these data and the smaller fit residuals
($R_w$).

The effective lattice constant ($a$) from \tabl{cubresults} is robust with
respect to the data reduction procedure.  This shows that the errors in the
correlation function introduced by the data reduction procedures are not severe
enough to result in a measurable peak shift. Thus, the geometric structure
information is readily accessible from the UEC data despite the uncontrolled
aspects of the data reduction protocol.

We note that the estimated lattice parameter is smaller than that for bulk
gold. This is not a physical effect, or a result of the data reduction
protocol, but rather due to a lack of instrument calibration.  Driven primarily
by the interests of differential structural changes, UEC frequently treats the
data using the known ground state structure as the reference.  In the case
where the ground state structure is not known, only relative changes can be
obtained.  Given the very small uncertainty of the refined lattice parameters
shown here, these ground state data could be used to calibrate the UEC setup by
comparing to data from the same particles measured with x-rays in a
conventional PDF measurement, thus making the absolute structural solution from
UEC possible.

\begin{table}[t]
\centering
\caption{
\label{table:cubresults}
Refinement parameters from cuboctahedral fits.  Here, $a$ is the effective
lattice parameter derived from nearest neighbor distances and $U_{iso}$ is the
isotropic ADP (see text). No standard deviations are shown on the refined
parameters as we don't know uncertainties reliably on the raw intensities.
}
\begin{tabular}{p{0.22\columnwidth}
p{0.22\columnwidth}p{0.22\columnwidth}p{0.22\columnwidth}}
    \hline
    \hline
            &   $R_w$       &   $a (\mAA)$      &   $U_{iso} (\mAA^2)$  \\
    \hline
    \bkgda  &   0.27        &   4.117           &   0.0250              \\
    \bkgdb  &   0.21        &   4.115           &   0.0264              \\
    \bkgdc  &   0.20        &   4.116           &   0.0167              \\
    \bkgdd  &   0.20        &   4.116           &   0.0185              \\
    \hline
    \hline

\end{tabular}
\end{table}

\section{Conclusions}

We have shown the equivalence of the mRDF commonly used in UEC and the PDF
typically employed in x-ray and neutron powder diffraction.  This has allowed
us to use powder diffraction modeling techniques to refine cuboctahedral,
icosahedral and decahedral models to the PDF derived from UEC data. We have
found that the cuboctahedral model agrees best with the data.  This
demonstrates that UEC data can be of sufficient quality to differentiate
between different nanoparticle structure models.

We have tested the cuboctahedral model with multiple PDFs produced using
different data reduction protocols that differ in how the intensity background
is removed and the data normalized.  Given the large differences in the
background curves fit through the raw intensity data, we are able to determine
the effective lattice parameter of the cuboctahedral model with remarkable
certainty.  We cannot determine with certainty the thermal displacement
parameter within the model because this parameter is strongly influenced by the
form of the intensity background.

The figure of merit of our best fits (cuboctahedral \bkgdc~and \bkgdd, $R_w =
0.20$) is comparable to results from high-quality nanoparticle x-ray PDF
studies.\cite{masad;prb07,gilbe;jac08}  The cuboctahedral model affords little
flexibility to compensate for systematic errors, so the results indicate that
these errors are not significantly worse than what one would expect from
synchrotron x-ray data. We conclude that, with careful data reduction and a
standard system for calibration, UEC is applicable as a quantitative
nanostructure probe.

\section{Acknowledgments}

This work was supported by the US National Science foundation through Grant
DMR-0703940.

\bibliographystyle{nar}
\bibliography{abb-billinge-group,everyone,billinge-group}

\end{document}